\documentstyle[rotate,epsf,floats,prb,aps]{revtex}

\begin{document}
\draft

\twocolumn[\hsize\textwidth\columnwidth\hsize\csname
@twocolumnfalse\endcsname

\title{Snaking of a fluid in two dimensions} \author{Maksim Skorobogatiy}
\address{Department of Physics, Massachusetts Institute of Technology,
Cambridge 02139, USA}

\maketitle

\begin{abstract}
The nature of physical processes can depend substantially on the
dimensionality of a system. One such example are buckling 
instabilities, which arise from the competition between axial 
compression and bending in elastic filaments. Thus, coiling of a jet of 
viscous fluid falling on a substrate (honey poured on a toast) 
\cite{maha1,griffiths} behaves quite differently from the coiling of 
viscous sheets (volcanous lava sliding on a crust) 
\cite{johnson,ramberg}. Here we consider a novel effect of coiling of a 
fluid jet confined to a two dimensional film.
\end{abstract}


\vskip2pc]

\narrowtext

\begin{figure}
\epsfxsize=3.0in
\epsfysize=3.0in
\centerline{\epsfbox{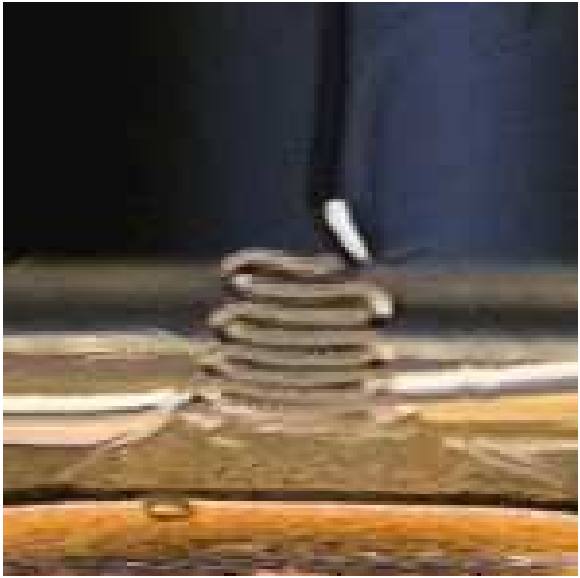}}
\vskip 0.2truein
\caption{
Snapshot of a coiling filament of liquid soap in a liquid soap film.
Experiments were made in the range of parameters: coiling frequency
3-11 Hz, filament radius 0.1-1 mm and a width of a coil 5-13 mm.
\label{fig1}} 
\end{figure}

While coiling in three dimensions is easy to observe by pouring of a 
viscous fluid on a substrate \cite{taylor}, a two dimensional analog of 
this process is not that straightforward as viscous sheets are usually 
unstable toward fluctuations leading to a breakup of a sheet. In our 
experiment we use a soap film stretched on a frame and a soap jet 
falling from the top of a frame thus confined to a
two dimensional surface of a film Fig. ~\ref{fig1}. We derive and check a 
scaling law that connects the filament radius $r$ to a coil size $L$.    

If the coiling frequency is low enough so that effects of 
inertia are unimportant $\omega \ll (\frac{g}{L})^{\frac{1}{2}}$, the 
coiling instability is determined by the competition between a 
torque exerted by the gravitational force and a torque due to the bending
of a filament. Assuming that the Young's modulus of soap is $E$ 
a torque due to the bending of an elastic filament with radius $r$ and 
curvature $L$ will be proportional to $E \frac{r^4}{L}$. A torque due to 
the gravitational force is $\rho g r^2  L^2$, where $\rho$ is the density of 
soap. Equating both torques gives 
\begin{equation}
L \sim (\frac{E}{\rho g})^{\frac{1}{3}} r^{\frac{2}{3}}
\label{one}
\end{equation}
which is a well known result for an elastic rod models 
\cite{maha2,smith,tchavdarov}.

\begin{figure}
\epsfxsize=3.0in
\epsfysize=3.0in
\centerline{\rotate[r]{\epsfbox{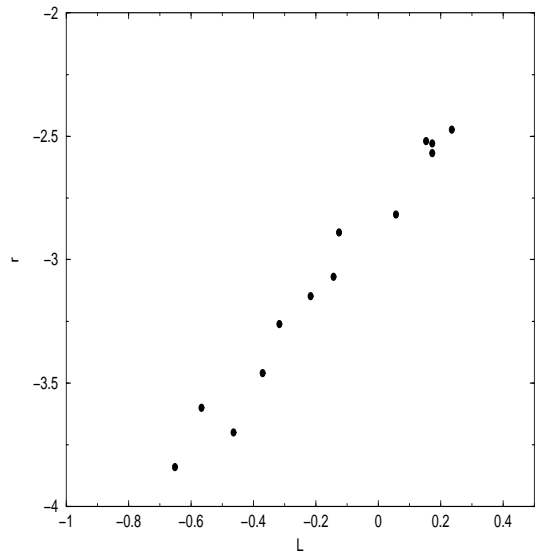}}}
\vskip 0.2truein
\caption{Log-log plot of the filament radius versus the width of a coil.
The data are fitted by the relation $L \sim r^{0.62 \pm 0.04}$ and is in a
good agreement with the theoretical estimate (\ref{one}).
\label{fig2}} 
\end{figure}

We studied coiling of liquid soap flowing from the top of a metal 
frame on which the soap film was stretched. Coiling was observed at the 
bottom of the frame for different frame heights thus 
establishing a dependence of coil size $L$ on a filament radius $r$. 
A camera with a 20 fold zoom was used to measure a filament radius and 
a coil width with an accuracy of $5\%$.
The data was taken in a regime where the radius of the filament was at 
least five times smaller than a coil size and a frequency was small 
enough to be detected within an accuracy of $10\%$ using a standard 
camera frame frequency. The data in Fig. ~\ref{fig2} collapses to the 
scaling law $L \sim r^{0.62 \pm 0.04}$ and agrees well with 
the theoretically predicted scaling.

While such relatively unimportant effects as air drag, non-Newtonian 
effects and relaxation of a viscous flow have been neglected it is 
surprising to find that the effects due to the viscosity and surface 
tension do not modify scaling law (\ref{one}) derived essentially for an
elastic rod model. Apparently because the thickness of a film is much 
smaller then the radius of a filament and an interface relaxation time 
due to the low surface tension is high, surface tension and viscosity do 
not modify elastic rod like scaling laws at low frequencies.

\end{document}